\documentclass[journal=nl,manuscript=article]{achemso}
\usepackage{amsmath}
\usepackage{amssymb}
\usepackage{graphicx}
\usepackage{epsfig}
\usepackage{dcolumn}
\usepackage{bm}
\usepackage{rotating}
\usepackage{multirow}
\usepackage{xspace}
\usepackage[table]{xcolor}
\usepackage[english]{babel}
\usepackage{colortbl}
\usepackage{lineno}

\def\ve{{\varepsilon}}

\def\bk{{\bf k}}
\def\bq{{\bf q}}

\def\br{{\bf r}}

\def\kprime{$\rm K^\prime$ }

\newcommand{\ket}[1]{ | #1 \rangle }
\newcommand{\bra}[1]{ \langle #1 | }

\author{Yiming Pan}
\affiliation{Institut f\"ur Theoretische Physik und Astrophysik, Christian-Albrechts-Universit\"at zu Kiel, Kiel, Germany} 
\author{Fabio Caruso}
\email{caruso@physik.uni-kiel.de}
\affiliation{Institut f\"ur Theoretische Physik und Astrophysik, Christian-Albrechts-Universit\"at zu Kiel, Kiel, Germany} 

\title{ Strain-induced activation of chiral-phonon emission in monolayer WS$\bf{_2}$ } 

\begin{document}	

    \begin{abstract}
    
We report a theoretical investigation of the ultrafast dynamics of electrons
and phonons in strained monolayer WS$_2$ following photoexcitation. We show
that strain substantially modifies the phase space for electron-phonon
scattering, unlocking new relaxation pathways that are unavailable in the
pristine monolayer. In particular, strain triggers a transition between
distinct dynamical regimes of the non-equilibrium lattice dynamics
characterized by the emission of chiral phonons under high strain and
linearly-polarized phonons under low strain. For valley-polarized electronic
excitations, this mechanism can be exploited to selectively activate the
emission of chiral phonons -- phonons carrying a net angular momentum.  Our
simulations are based on state-of-the-art ab-initio methods and focus
exclusively on realistic excitation and strain conditions that have already
been achieved in recent experimental studies. Overall, strain emerges as a
powerful tool for controlling chiral phonons emission and relaxation pathways
in multivalley quantum materials. 
\end{abstract}
	
    \maketitle
\section{Introduction}
Strain is a powerful platform for tailoring the emergent properties of
condensed matter, with the capability of disclosing new phases and quantum
phenomena unavailable in their unaltered states. Several remarkable instances
of this paradigm have recently been demonstrated including, for example,
strain-induced  quantum paraelectric to ferroelectric phase
transitions,\cite{haeni2004room} the stabilization of
superconductivity,\cite{ruf2021strain} reversible magnetic phase
transitions,\cite{cenker2022reversible} and topological phase
transitions.\cite{lin2021visualization}  Additionally, strain engineering has
become a pivotal ingredient in the semiconductor manufacturing industry as a
tool to enhance electron mobilities in
transistors.\cite{ChuSun2009,Ponce_GaN_2019} 

Two dimensional materials (2D), and particularly transition metal
dichalcogenides (TMDs) in their monolayer form, are  particularly suitable for the
exploration of novel strain-induced phenomena owing to the interplay of
flexibility and structural stability, enabling to reach strain regimes
inaccessible in the bulk. TMDs can withstand strain up to 10\% before
undergoing structural damage.\cite{dBertolazzi2011,peng2020strain}
Additionally, high-strain regimes have been reached through a variety of
techniques, including substrate lattice mismatching,\cite{Chae2017,
peng2020strain} substrate thermal expansion,\cite{Plechinger_2015,
ahn2017strain} van der Waals coupling,\cite{ZhaoDing2018} and proton
irradiation.\cite{Blundo2020}

Strain can be exploited to fine-tune specific features in the electron band
structures,\cite{feng2012strain,Fang_Carr_2018,Schmidt_2016} providing a route
to directly influence optical response, scattering channels, and transport
properties of TMD monolayers. In particular, strain alters the energy level
alignment between the K to the Q($\Sigma$) high-symmetry points in the
conduction band.\cite{Sohier2019, Sohier_2023,SGCPM2019, Yuan2016,Pimenta2022}
This effect induces a transition from direct to indirect band gap under
compressive
strain,\cite{ShiPan2013,SongYang2017,WangKutana2014,wang2015strain,Blundo2020,Desai2014}
with distinctive fingerprints in photoluminescence
measurements.\cite{Niehues2018, ZhuWang2013,
PakLee2017,HePoole2013,CastellanosAndres2013} Conversely, tensile strain
increases the energy separation between the K and Q high-symmetry points in the
conduction band, thereby, suppressing intervalley scattering and resulting in
the increase of carrier mobility in transport
measurements.\cite{ShenPenumatcha2016,Quereda_2017,Datye2022,
ZhangChengLiu2020,Ghorbani-Asl2013}

More generally, the impact of strain on valley degrees of freedom in TMD
monolayers emerges as a promising route to engineer physical phenomena that
directly result from the multivalley character of the electron band structure.
\cite{Ge2013,Piatti2018,Piatti_2019,ZhangChengLiu2020,SGCPM2019}
Valley-dependent optical selection rules -- arising from the coexistence of
three-fold rotational invariance and a non-centrosymmetric crystal structure --
allow to selectively inject carriers at either the K or K$'$ high-symmetry
points via absorption of circularly-polarized photons, enabling the
establishment of a finite valley polarization (namely, an anisotropic
population of the K and K$'$ valleys) and the exploitation of valley degrees of
freedom.\cite{MoS2_valleytronic,Cao2012}  This effect serves as the foundation
for a variety of physical phenomena that are unique to TMD monolayers, 
including the emergence of chiral valley excitons and the valley Zeeman
effect, \cite{valley_exciton,Srivastava2015} circular
dichroism,\cite{Pan2023} chiral phonon
emission,\cite{chiral_phonon_experiment,LiWangJin2019,chiral_phonon_theory,ChenWangFeng2023}
as well as different flavours of Hall effects.\cite{MoS2_valleytronic,YaoWangXiao2008,MakMcGrill2014}  
 In particular, chiral phonon emission can underpin a variety of novel physical phenomena in TMD monolayers. The finite angular momentum associated with chiral phonons imparts a finite magnetic moment to the lattice, if the ions carry nonzero Born effective charges.\cite{Juraschek2017, Nova2017, Xiong2022, Tauchert2022} This magnetic moment enables a direct coupling of the lattice to external magnetic fields, giving rise to phonon Zeeman effects,\cite{Baydin2022, Juraschek2017} which are closely analogous to their electronic counterparts.\cite{valley_exciton,Srivastava2015} Chiral phonons have further been proposed as a route towards nontrivial topological properties of the lattice, such as the phonon Hall effect.\cite{chiral_phonon_theory, phonon_hall}
Additionally, ultrafast
phenomena in TMDs are governed by electron-phonon scattering processes
involving inter- and intra-valley transitions accompanied by the emissions or
absorption of a phonon.\cite{MoS2_TDBE, Perfetto2023,dfsct_WSe2, MX2_eph_soc,
MX2_eph, Song2013transport, MoS2_UDES, WangMeng2023} These processes are
extremely sensitive to the phase space available for phonon-assisted
transitions, and are thus likely to be largely affected by the strain.
Overall, these considerations suggest opportunities to directly influence
intervalley scattering processes and the ensuing nonequilibrium dynamics of
phonons and charge carrier using strain. Yet, these phenomena remain mostly
unexplored out of equilibrium.

\begin{figure*}[t]
		\begin{center}
			\includegraphics[width=0.7\textwidth]{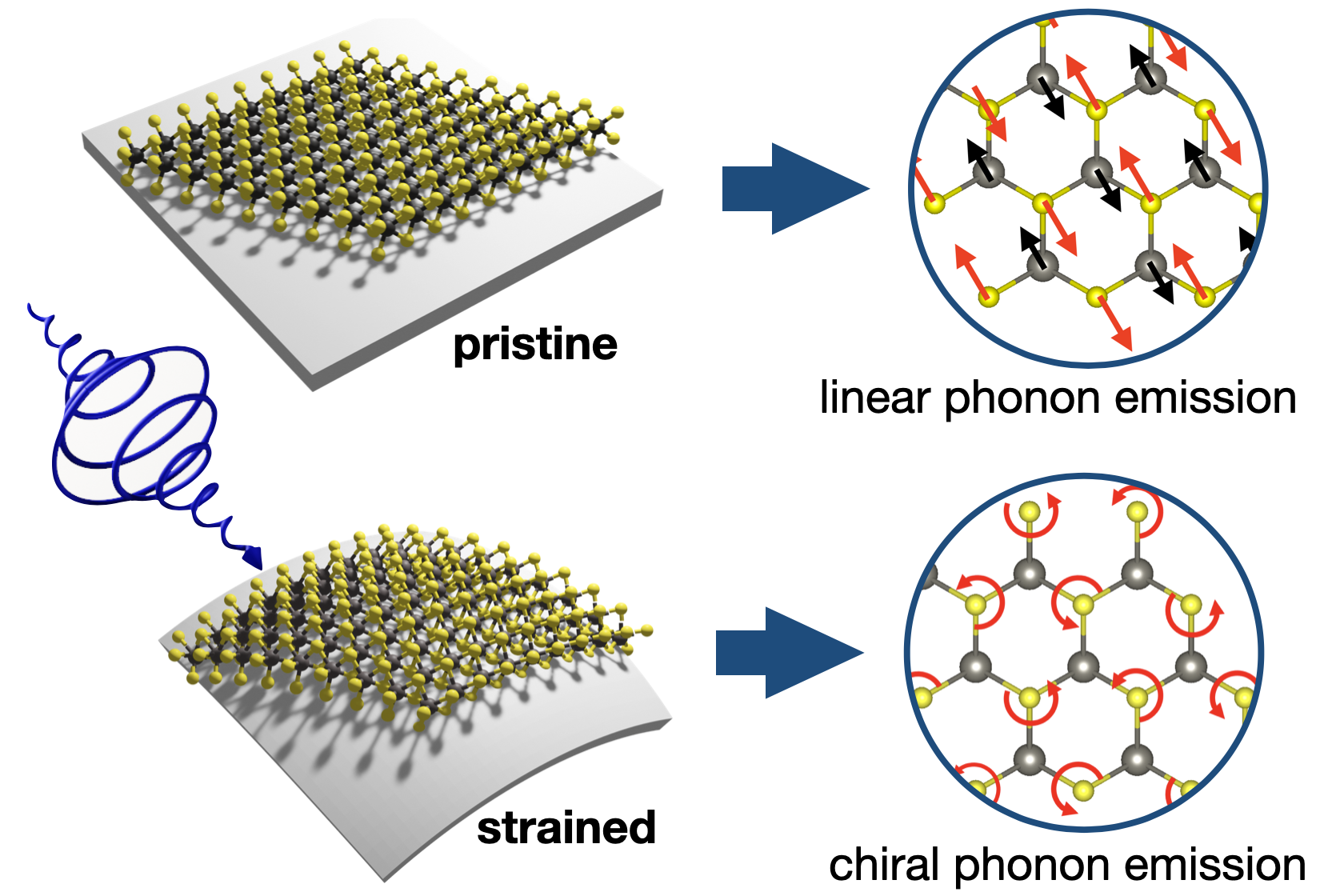}	
		\end{center}
		\caption{Schematic illustration of the influence of strain on the phonon emission in monolayer WS$_2$. }
\label{sketch} 
	\end{figure*}

In this work, we explore strain-induced modifications of the ultrafast electron
and phonon dynamics in monolayer WS$_2$ via state-of-the-art ab-initio methods.
Our investigations are based on the time-dependent Boltzmann equation (TDBE)
and concentrate on realistic strain conditions that have already been achieved
in experiments.  We demonstrate that strain influences profoundly the phase
space available for electron-phonon scattering, altering the accessible
scattering channels for the relaxation of hot carriers.  Specifically, for the
case of valley-polarized electronic excitations strain can be exploited to
switch among qualitatively different dynamical regimes, characterized by the
emission of linearly polarized phonons at low strain and circularly polarized
(chiral) phonons at high strain, as depicted in Fig.~\ref{sketch}.  Overall, these findings demonstrate a
powerful route to
 enhance the chiral phonon emission in multivalley TMDs
and open up new opportunities to directly tailor the chirality
of lattice vibrations in photo-driven solids.

 \section{Results and Discussion}
\subsection{Influence of strain on electrons and phonons at equilibrium}
\begin{figure*}[t]
		\begin{center}
			\includegraphics[width=1.0\textwidth]{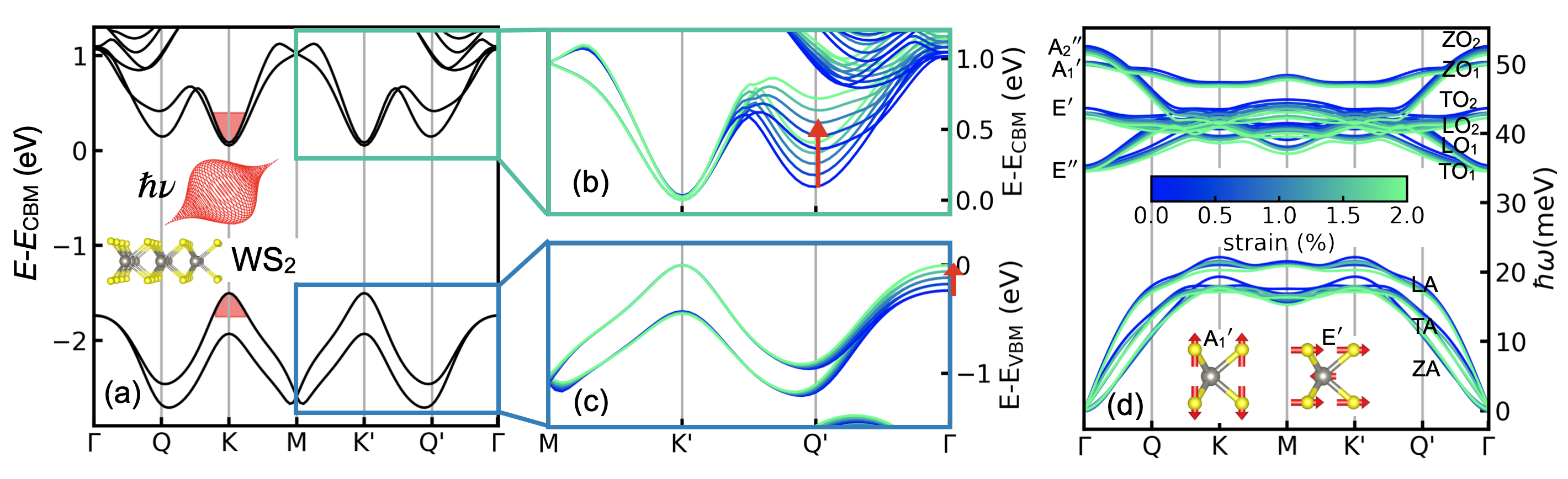}	
		\end{center}
		\caption{Influence of strain on the electron band structure and
phonon dispersion of monolayer WS$_2$. (a) Electronic band structures along the
high symmetry path $\Gamma$-K-M-K$'$-$\Gamma$, the red shading indicates the
carrier population induced by a circularly polarized light pulse. Conduction
(b) and valence bands (c) along the high symmetry path M-K-$\Gamma$, with
energies relative to the CBM and VBM, respectively. The color coding reflects
strain values ranging between 0 and 2~\%. (d) Influence of strain on the
phonon dispersion. Inset: Vibrations of motifs corresponding to the $A_1^\prime$ and $E^\prime$ modes.} \label{bnd_ph} 
	\end{figure*}

    The band structure of monolayer WS$_2$ as obtained from density functional
theory (DFT) is reported in Fig.~\ref{bnd_ph}~(a). To illustrate the effects of
tensile strain on the electronic properties and phonon dispersion of monolayer
WS$_2$, we perform DFT and density functional perturbation theory (DFPT)
calculations for an increase of the in-plane lattice constant ranging from 0\%
to 2\% with an increment of 0.5\%. 
The phonon symmetry analysis is reported in supplementary Sec.~S1 and Fig.~S1.
In Figs.~\ref{bnd_ph}~(b) and (c), we
illustrate the influence of different strain conditions on the conduction and
valence bands, respectively.  Energies in the conduction (valence) band are
relative to the CBM (VBM).  The effect of strain manifests itself primarily in
the conduction band.  With increasing strain the spin-split Q valley -- located
at the midpoint between the $\Gamma$ and K/\kprime high-symmetry points in the
conduction bands -- is blue-shifted to higher energies, leading to an increase
in the energy separation between the Q point and the CBM. Conversely, the
spin-degenerate $\Gamma$ valley in the valence bands follows the opposite
trend.  Tensile strain reduces the energy difference between the K and $\Gamma$
high-symmetry points and for strain levels beyond 2\% monolayer WS$_2$
undergoes a transition from direct to indirect band-gap semiconductor, as
demonstrated by recent photoluminescence experiments.\cite{Blundo2020} The
origin of the band structure dependence on strain has been discussed
extensively in earlier studies and it can be attributed to the reduction of
monolayer thickness when strain is
applied.\cite{Sayers2023,Trovatello2020,Sohier2019} 	
{While we focus here on the case of bi-axial strain, uni-axial strain induces similar band structural renormalization as illustrated in Fig.~S5  and discussed in Sec.~S2.} 

The phonon dispersion for different values of biaxial strain is reported in
Fig.~\ref{bnd_ph}~(d). In general, strain induces a softening of the phonons
which arises from the weakening of bonds as the interatomic distances are
increased. The phonon frequencies, however, are not homogeneously softened for
different modes. For the $E^\prime$ mode we observe a softening by 0.87 meV per
\% of strain. Conversely for the $A_1^\prime$ mode, the softening amounts to
less than 0.1 meV per \% of strain. This result agrees well with
strain-dependent Raman spectroscopy experiments \cite{Rice2013, Conley2013} and
the analysis of mode Gr\"uneisen parameters.\cite{Huang2014,Sevik2014} The
different dependence of $E^\prime$ and $A_1^\prime$ phonon energies on strain
can be ascribed to the anharmonicity of atomic potentials and the increase of
in-plane atomic distances. As a result, the $E^\prime$ mode with in-plane
vibrations are strongly softened compared to the $A_1^\prime$  phonon which
vibrates in the out-of-plane direction.  The vibrations of these  modes are schematically illustrated in the inset of Fig.~\ref{bnd_ph} (d).

Having discussed the effects of strain on the electronic and vibrational
properties under equilibrium conditions, we proceed next to inspect to which
extent strain can influence the ultrafast dynamics of electrons and phonons out
of equilibrium.  Overall, while the minor changes of the phonon dispersion
under strain are likely inconsequential for the ultrafast dynamics, the
substantial strain-induced band-structure modifications are expected to
profoundly influence the phase space available for phonon-assisted electronic
transitions. In particular, in the following we show that the renormalization
of band energy at the Q point can lead to the selective activation/deactivation
of additional scattering channels, leading to qualitatively different regimes
for the non-equilibrium lattice dynamics. 

\subsection{Coupled electron-phonon dynamics in strained WS$_2$}  
We focus below on the case of valley-selective circular dichroism and on the
ensuing non-equilibrium dynamics of electrons and phonons.  The interplay of
threefold rotational invariance and the lack of inversion symmetry in hexagonal
monolayer TMDs is responsible for distinctive valley-selective  optical
selection rules.  In particular,  photo-excitation with circularly-polarized
photons and photon energies in vicinity of the fundamental gap can lead to the
formation of electron-hole pairs located at either the K or \kprime
high-symmetry points depending on the light helicity.  Pulses shorter than
100~fs have been shown to generate (valley-polarized) excited carrier density
of the order of 10$^{12}$ to 10$^{13}$ cm$^{-2}$ in
TMDs\cite{Wallauer2021,Trovatello2022,Trovatello2020_2,Perfetto2023}. 

To investigate the nonequilibrium dynamics of electrons and phonons, we solve
the TDBE in the time domain by accounting explicitly for electron-phonon and
phonon-phonon scattering processes.  Within the TDBE approach, the evolution of
time-dependent electron distribution function $f_{n \bf k}$ and phonon number
$n_{\bf q\nu}$ is described  by the following set of coupled first-order
integro-differential equations:
    \begin{align}
    \partial_t f_{n\bk}(t) &= \Gamma^{\rm ep}[f_{n\bk}(t),n_{\bq \nu}(t)] \label{TDBE1} \\
    \partial_t n_{\bq \nu}(t) &= \Gamma^{\rm pe} [f_{n\bk}(t),n_{\bq \nu}(t)] + \Gamma^{\rm pp}[n_{\bq \nu}(t)] \label{TDBE2}
    \end{align} 
with $\partial_t = \partial / \partial t$.  Here, $\Gamma^{\rm ep}$,
$\Gamma^{\rm pe}$, and $\Gamma^{\rm pp}$ denote the collision integrals due to
electron-phonon, phonon-electron (ph-e), and phonon-phonon (ph-ph)
interactions, respectively, for which explicit expressions have been reported
elsewhere \cite{TDBE,TDBE2,TDBE4,MoS2_TDBE,GirottoCaruso2023,Pan2023}.
Radiative recombination is neglected hereafter as it takes place on nanosecond
timescales which are much longer the characteristic time of electron-phonon
scattering.  Additionally, intervalley electron-electron scattering processes
(K$\rightarrow$\kprime or viceversa) do not influence the valley depolarization
dynamics owing to momentum conservation.  These effects are thus neglected in
our simulations.  We further neglect electron-hole interaction. This choice is
compatible with the large carrier densities considered in our work and in
earlier pump-probe investigations ($>10^{13}$~cm$^{-2}$).  Recent
photoluminescence studies indicate that at these carrier densities
electron-hole interactions are screened, suppressing the formation of excitons,
leading to formation of an electron-hole plasma.\cite{YuBataller2019} We solved
Eqs.~\eqref{TDBE1} and \eqref{TDBE2} fully from first principles based on our
recent implementation of the TDBE approach within the {\tt EPW}
code\cite{EPW2023}.  The electron-phonon coupling matrix elements are obtained
from DFT and DFPT with norm-conserving Vanderbilt (ONCV)
pseudopotentials\cite{ONCV} and Perdew-Burke-Ernzerhof generalized gradient
approximation (GGA-PBE) to the exchange-correlation,\cite{GGA_PBE} as
implemented in {\tt Quantum ESPRESSO} \cite{Quantum_epsresso,Baroni2001} and
interpolated using maximally-localized Wannier functions within {\tt EPW} using
the {\tt Wannier90} code as a library.\cite{Giustino2007,pizzi2020wannier90}
Equations~\eqref{TDBE1} and \eqref{TDBE2} are solved as an initial value
problem by discretizing the time derivative using Heun's method and
recalculating the collision integrals at each time step on homogeneous
Monkhorst-Pack grids consisting of 120$\times$120$\times$1 ${\bf k}$ and ${\bf
q}$  points.  We used a time step of 1 fs and considered a total simulation
time of 6 ps.  All computational details are reported in Supporting
Information.  
\begin{figure*}[t]
		\begin{center}
			\includegraphics[width=1.0\textwidth]{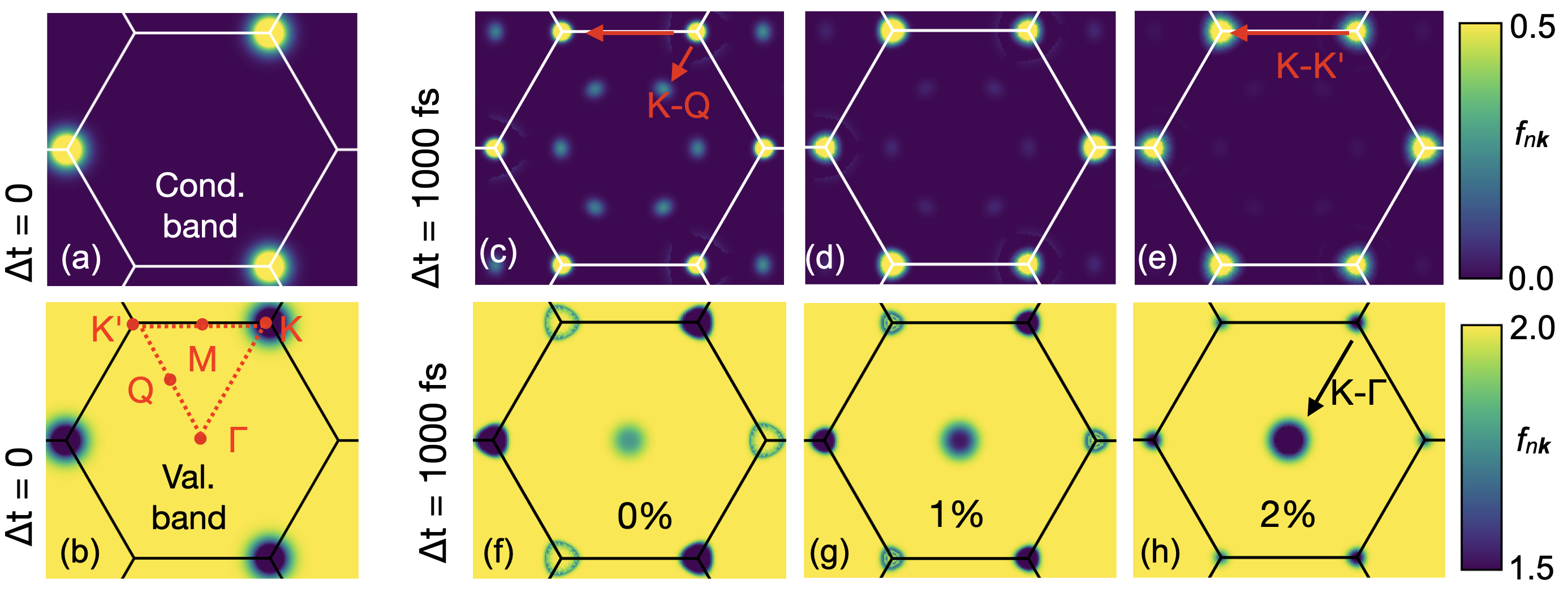}	
		\end{center}
		\caption{Influence of strain on the valley-depolarization
dynamics of electrons and holes in monolayer WS$_2$. (a-b) Momentum-resolved
electron distribution in the conduction and valence bands -- generated by
ultrashort circularly-polarized light pulses resonant with the band gap at the K
point --  for a carrier density of $2.0\times 10^{13} {\rm cm}^{-2}$.  The
electron populations $f_{n\bf k}$ are summed over both spin channels for the lowest conduction and highest valence bands. (c-e)  
Momentum-resolved electron distribution in the conduction bands under 0\%, 1\% and 2\%
biaxial strain for a time delay of 1~ps after photo excitation. (f-h)
Momentum-resolved electron distribution in the valence bands for a 
time delay of 1~ps for different values of strain. }
\label{electron_dynamics} 
	\end{figure*}      
    
As initial state, we consider electron and phonon distribution functions that
mimic the conditions established following absorption of a circularly-polarized
light pulse. The lattice is initially at thermal equilibrium for $t=0$, with
phonon occupations defined according to Bose-Einstein statistic at room
temperature (300~K).  For momenta within the hexagonal BZ, the initial electron
distributions $f_{n{\bf k}}(t=0)$ in the conduction and valence bands are
illustrated in Figs.~\ref{electron_dynamics}~(a) and (b), respectively.
$f_{n{\bf k}}(t=0)$ is defined by considering a photoexcited electron (hole)
density of 2.0$\times 10^{13}$ cm$^{-2}$ at the K valley in the conduction
(valence) band, whereas electron and hole occupations at K$^\prime$ are
unchanged with respect to thermal equilibrium. 

\begin{figure*}[t]
		\begin{center}
        \includegraphics[width=1.0\textwidth]{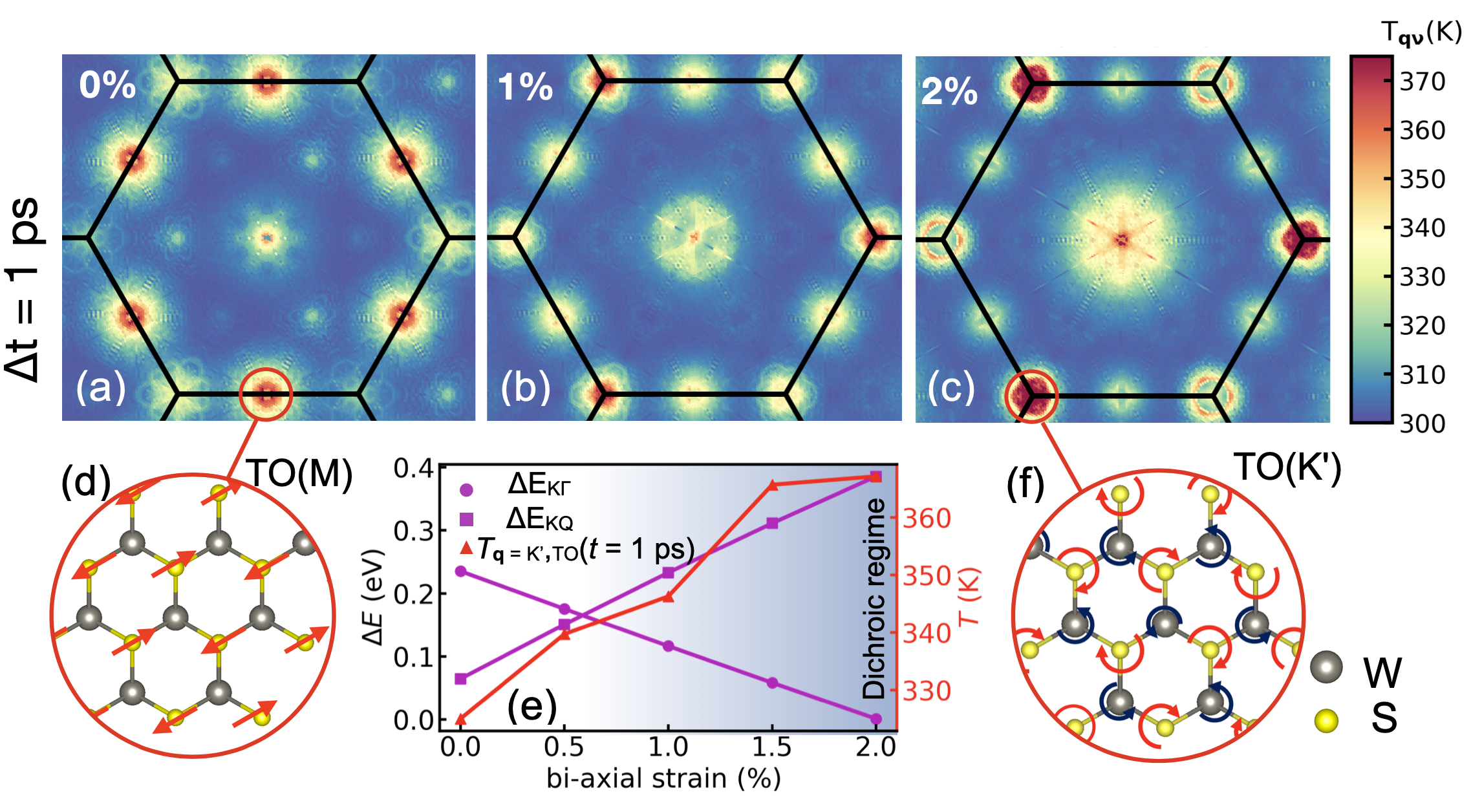}	
		\end{center}
		
		\caption{Nonequilibrium phonon dynamics in strained monolayer
WS$_2$ due to electron-phonon  and phonon-phonon interactions. (a-c) Momentum-resolved effective
vibrational temperature $T_{\bq, \rm TO}$ for the strongly-coupled TO mode 
for crystal momenta in the Brillouin zone for a time delay of $1$~ps and different levels of strain. (d)
Schematic representation of the vibrational motif arising from linearly-polarized phonons at the M 
high-symmetry point. (e) Energy difference between the K and 
Q ($\Gamma$) valleys in the conduction (valence) bands as a function of strain. 
The $y$-axis marks the effective vibrational temperature $T_{\bq, \rm TO}$ obtained at 
the \kprime valley under different levels of strain for 1~ps time delay. 
(f) Schematic representation of the strongly-coupled chiral TO phonons at the 
\kprime high-symmetry  point. }
    \label{phonon_dynamics} 
	\end{figure*}

Figures~\ref{electron_dynamics}~(c)-(e) illustrate the electron distribution
function in the conduction band after 1~ps following photoexcitation for 0, 1,
and 2\% strain, respectively.  The valley polarization of carriers in the
conduction bands, reflected by the anisotropy in the population of the K and
\kprime valley, is found to decays within 1 ps for all values of strain.
Remarkably, strain directly influences the decay path of excited carriers.  In
unstrained WS$_2$, electrons at K can directly scatter to either the \kprime or
Q valley by emitting phonons.  As strain increases, however, the scattering of
electrons from K to  Q is gradually suppressed, as reflected by the decrease of
electron occupation at the Q high-symmetry point in
Fig.~\ref{electron_dynamics}~(d) and (e).  This behaviour can be attributed to
the increase in the energy separation  $\Delta E_{KQ} \equiv -(E_K - E_Q)$
between the K and Q valleys in the conduction band.  $\Delta E_{KQ}$ increases
linearly from 60~meV for the unstrained monolayer up to 390~meV for 2\% strain,
as illustrated in Fig.~\ref{phonon_dynamics}~(e).  For phonon-assisted
transitions, energy conservation requires $\ve_{\rm fin} - \ve_{\rm in} \simeq
\hbar\omega_{\bq\nu}$, where $\ve$ denote the initial and final electron energy
and $\hbar\omega_{\bf q \nu} < 50$~meV is the phonon energy. This condition
cannot be satisfied for strain values larger than 1\% leading to the
deactivation of the K$\rightarrow$Q scattering channel.  This result is also
consistent with the increase of electron mobility in strained monolayer
WS$_2$.\cite{ZhangChengLiu2020} 

The electron distribution in the valence bands  1~ps after excitation is shown
in Figs.~\ref{electron_dynamics}~(f)-(h) for 0, 1, and 2\% strain,
respectively.  Also in the valence bands strain influences profoundly the
carrier dynamics by altering the accessible scattering channels.  In unstrained
monolayer, K$\rightarrow$\kprime transitions are spin forbidden, whereas
K$\rightarrow\Gamma$ transitions are energy forbidden.  Correspondingly, the
valley polarization is preserved beyond 1~ps, as shown in
Fig.~\ref{electron_dynamics}~(f).  In presence of strain, the energy difference
$\Delta E_{K\Gamma}$ between K and $\Gamma$ valley decreases from 240~meV to
about 1~meV at 2\%, as shown in Fig.~\ref{phonon_dynamics}~(e), activating the
K$\rightarrow\Gamma$ scattering channel and substantially accelerating the
electron dynamics in valence bands [Figs.~\ref{electron_dynamics}~(g)-(h)].  

Recent theoretical and experimental studies revealed that the non-equilibrium
phonon population established following photoexcitation are extremely sensitive
to the phase space available for phonon-assisted electronic
transitions.\cite{MoS2_TDBE,bP_UEDS} These considerations, alongside with the
findings discussed above, indicate that strain could serve as a means to
selectively influence the mechanisms of phonon emissions and the ensuing phonon
populations in monolayer WS$_2$ and related compounds.  In the following, we
proceed to corroborate this hypothesis by quantifying the non-equilibrium
phonon population established upon thermalization of valley-polarized electrons
and holes. 

As an indicator for the nonequilibrium phonon populations established as a
result of the coupled electron-phonon dynamics, we consider below the mode
resolved  effective phonon temperature\cite{MoS2_TDBE} $T_{\bq \nu} =\hbar \omega_{\bq\nu}
[{k_{\rm B}{\rm ln}(1+n_{\bq \nu}^{-1})}]^{-1}$,   where $k_{\rm B}$ is the
Boltzmann constant, $\omega_{\bq\nu}$ the phonon frequency, and $n_{\bq \nu}$
is the phonon population obtained from Eq.~\eqref{TDBE2}.  In short, $T_{\bq
\nu}$  is directly related to the phonon population, it is constant  at thermal
equilibrium, whereas it may differ for distinct vibrational modes out of
equilibrium. In particular, the quantity $T_{\bq \nu}$ enable to directly
monitor transient changes of phonon population with full momentum resolution.
The {nonequilibrium phonon} temperature {for the strongly coupled TO branch}
established after 1~ps --  reported in Figs.~\ref{phonon_dynamics}~(a-c) for 0,
1, and 2\% strain, respectively -- indicates that strain profoundly influences
the phonon populations excited throughout the valley depolarization dynamics.  {The strain effects induce similar changes to other phonon branches, as demonstrated in Fig.~S2 and Fig.~S3. \cite{supp}}
In the absence of strain, the  effective phonon temperature [Fig.~\ref{phonon_dynamics}~(a)]
reflects an enhancement of the phonon population for crystal momenta $\bq$ in
the vicinity of the M, Q, and $\Gamma$ high-symmetry points.  While $\Gamma$
phonons are emitted throughout intravalley relaxation (and thus inconsequential
for the valley depolarization), the emission of M and Q phonons can only result
from electronic transitions mediated by the Q valley owing to momentum
conservation.  These results indicate that the most likely scattering pathway
for valley-polarized carriers in pristine WS$_2$ involves a Q-mediated two-step
process ($\rm  K \rightarrow Q \rightarrow K^{\prime}$),  whereas the direct
transition ($\rm  K \rightarrow K^{\prime}$) occurs at a significantly lower
rate.  

As lattice strain is introduced  [Figs.~\ref{phonon_dynamics}~(b) and (c)], the
emission of M and Q phonons is gradually suppressed, and for 2\% strain the
nonequilibrium phonon population is almost exclusively characterized by hot
phonons  at the \kprime and $\Gamma$ points.  Figure~\ref{phonon_dynamics}~(e)
further indicates that with the strain-induced increase of $\Delta E_{\rm KQ}$
and the ensuing suppression of K$\rightarrow$Q  scattering processes, the
temperature of \kprime phonons (right axis) increases linearly with strain,
saturating at about 2\% strain.  In other words, strain effectively deactivate
the $\rm  K \rightarrow Q \rightarrow K^{\prime}$ scattering channel, boosting
the emission of \kprime phonons.  Overall, these findings reveal the
possibility to exploit strain to selectively switch between two distinct
dynamical regimes: at low strain, the electron and phonon dynamics is governed
by two-step scattering processes $\rm  K \rightarrow Q \rightarrow K^{\prime}$
resulting in the emission of M and Q phonons; beyond a critical strain
threshold of about 1\%, conversely, the scattering pathways involve $\rm  K
\rightarrow K^{\prime}$ transitions accompanied by the emission of K$^\prime$
phonons. This latter regime is characterized by valley-polarized phonon
populations (with $T_{\rm  K, TO} \neq T_{\rm K^\prime, TO}$) which can persist
for several picoseconds (Fig.~S6), and that constitute  the vibrational counterpart of
the valleytronic paradigm in 2D materials.\cite{MoS2_valleytronic}
The full time dependence of the phonon population at the high-symmetry points in the BZ is illustrated in Fig.~S4.
Additionally, valley-polarized phonons have been predicted to give rise to
vibrational circular dichroism in ultrafast diffuse scattering
experiments,\cite{Pan2023}  which could provide a direct route for the experimental validation of these predictions. 

The crossover from M-phonon to K$^\prime$-phonon regimes also indicates a
transition from linearly-polarized phonon emission to chiral phonon emission.
The phonon chirality can be characterized by introducing phonon angular
momentum $L_{\bq \nu}=\bra{\epsilon_{\bq\nu}}\hat{\br}\times\hat{\bf
p}\ket{\epsilon_{\bq\nu}} $, obtained by projecting the phonon eigenvector on a
circularly polarized basis.  In short, $L_{\bq \nu} > 0$ ($L_{\bq \nu} < 0$)
corresponds to chiral phonon with right-handed (left-handed) circular
polarization, whereas non-chiral phonons are characterized by $L_{\bq \nu}=0$.
For strained WS$_2$, the phonon emission is governed by the strongly-coupled
TO(${\rm K^{\prime}}$) mode, which is characterized by $L_{\bq \nu}^z > 0$ and
by a distinctive circular polarization, as illustrated in
Fig.~\ref{phonon_dynamics}~(f).  The chiral character of the emitted phonons by electronic transitions among the K, K’, and $\Gamma$ high-symmetry points can further be corroborated by a group-theoretical analysis, as discussed in details in previous works.\cite{Song2013transport,TMD_symmetry2,He2020-xf} At the M point, all phonons
display linearly-polarized vibrations, as required by time-reversal symmetry
(TRS) which enforces the identity $L_{-\bq \nu} = -L_{\bq \nu}=0$ for all
TRS-invariant points, including $M$.  The vibration motif of the
strongly-coupled M phonons is illustrated in Fig.~\ref{phonon_dynamics}~(d).
Consequently, the angular momentum lost by valley polarized electrons is
primarily carried by phonon at \kprime in strained WS$_2$, whereas in
unstrained materials the angular momentum of phonon is distributed throughout
all modes in the BZ. 

\subsection{Conclusions}

In conclusion, we presented an ab-initio investigation of the ultrafast
electron and phonon dynamics in strained WS$_2$ monolayers. Our study
demonstrates the possibility to exploit strain as a route to selectively
control the relaxation pathways of photoexcited carriers. Remarkably, even
strain values as low as 1\% have the capacity to significantly alter available
scattering channels and relaxation pathways in the valence and conduction
bands. This has profound implications for the lattice dynamics, and it can lead
to qualitatively different nonequilibrium phonon populations. Specifically, in
the context of valley-polarized electronic excitations -- as induced, e.g., by
the absorption of circularly-polarized photons --, strain can trigger a
transition between distinct dynamical regimes of the lattice characterized by
the emission of chiral phonons under high strain and linearly-polarized phonons
under low strain. 

More generally, multivalley materials\cite{SGCPM2019} emerge as a powerful
platform to exert control over nonequilibrium  phenomena through the
manipulation of the electron-phonon scattering phase space via strain.  It is
worth noting that the strain conditions considered in our study have already
been achieved in experiments, thus the experimental validation of these
findings can readily be attempted.   
Specifically, the excitation of chiral phonon is expected to give rise to distinctive fingerprints in time-resolved diffuse scattering experiments, leading to vibrational circular dichroism in the scattering intensities upon excitation with the circularly polarized light.  Additionally, we estimate that the lattice remains in a non-thermal state characterized by chiral phonon populations persisting for 10 ps or beyond, namely, over timescales that are  within reach of time-resolved scattering experiments.\cite{Pan2023} 
 
 \section{Methods}

\subsection{ Computational details}

The ground-state properties of the monolayer WS$_2$ are calculated from density-functional theory (DFT) within the plane-wave pseudopotential method as implemented in the {\tt Quantum Espresso} code.\cite{Quantum_epsresso} We used fully-relativistic norm-conserving Vanderbilt (ONCV) pseudopotentials\cite{ONCV} and Perdew-Burke-Ernzerhof generalized gradient approximation (GGA-PBE) to the exchange-correlation functional.\cite{GGA_PBE} We used the DFT-relaxed crystal structure, a plane-wave kinetic energy cutoff of 120~Ry and a Brillouin zone sampling of $20\times20\times 1$ points. The phonon dispersion was calculated based on density-functional perturbation theory (DFPT) on a $6\times 6\times 1$ Monkhorst-Pack grid.\cite{Baroni2001} Electron band structure and phonon dispersion are interpolated onto dense grids consisting of $120 \times 120 \times 1$ $\bf k$ and $\bf q$ points via maximally-localized Wannier functions, as implemented in {\tt Wannier90}.\cite{pizzi2020wannier90}  Electron-phonon matrix elements are computed with the {\tt EPW} code.\cite{EPW2023,Giustino2007}  The inclusion of phonon-phonon interactions requires the third-order force constants, which we obtained with finite displacement methods by generating displaced patterns on the $4\times 4 \times 1$ super-cell with {\tt thirdorder.py} from the {\tt ShengBTE} module.\cite{LI20141747} Spin-orbit coupling is included at all levels of calculations.
The ultrafast electron-phonon dynamics is investigated via solution of the time-dependent Boltzmann Equation (TDBE) --  which we implemented in the {\tt EPW} code -- for both electrons and phonons in presence of electron-phonon and phonon-phonon scattering.  
A detailed overview of this approach and our implementation is provided in earlier works.\cite{TDBE,TDBE2,TDBE4,MoS2_TDBE} 

\subsection{Initial conditions of the ultrafast dynamics simulations} 
We consider light pulses resonant with the band gap of WS$_2$, populating the highest valence band (VB) and second lowest conduction band (CB+1). This choice is assuming spin-conserving  photoexcitation conditions.
To simulate the initial carrier distributions excited by circular polarized light, we use a Gaussian function centered at the K point to describe electron-hole excitations: $f({\bf k}) = {\rm exp}\{-\frac{||{\bf k}-{\rm K}||^2}{\eta^2}\}$. Correspondingly, the initial distribution is defined by $f_{\rm VB} ({\bf k}) = 1 - f({\bf k})$ and $f_{\rm CB+1} ({\bf k}) =  f({\bf k})$, while leaving other bands unexcited. 
The parameter $\eta$ is related to the excited carrier density via the condition $n = \frac{1}{N_{\bf k}\Omega} \sum_{c{\bf k}} f_{c{\bf k}}$, where $N_{\bf k}$ and $\Omega$ are respectively the number of $\bf k$ points sampled ($120 \times 120 \times 1$) and the area of the two-dimensional (2D) unit cell. The carrier density of $2\times 10^{13}~\rm cm^{-2}$ considered in this work is equivalent to around 0.017 electrons (holes) per formula unit, and the nonadiabatic phonon renormalization can be neglected according to previous studies\cite{GirottoCaruso2023,MariniCalandra2021}. Hence the assumption of fixed electronic and lattice band structure used in TDBE is justified. 
The initial phonon distributions are considered to be thermally distributed with Bose-Einstein function at 300 K.

\section{Data Availability}
The datasets are available from the authors upon reasonable request.
\section{Code Availability}
The TDBE code is available from the authors upon reasonable request.

\section{Acknowledgments} 
	This project has been funded by the
	Deutsche Forschungsgemeinschaft (DFG) -- project numbers 443988403. The authors gratefully acknowledge the computing time provided to them on the high-performance computer Lichtenberg at the NHR Centers NHR4CES at TU Darmstadt. The computation time is granted under the Project: p0021280.
    \section{Authors contributions} 
YP performed simulations and data analysis. FC conceived the research.

\section{Competing Interests}
The authors declare no competing interests.

	\bibliography{references}
\end{document}